\documentclass[iop]{emulateapj}

\usepackage{ulem}
\usepackage{bm}
\usepackage{color}

\newcommand\kms{\ifmmode{\rm km\thinspace s^{-1}}\else km\thinspace s$^{-1}$\fi}
\newcommand\gaia{{\it Gaia\/}}
\newcommand\hip{{\it Hipparcos\/}}
\newcommand\estar{$\eta$~Gem}
\newcommand\jaavso{Journal of the American Association of Variable Star Observers}

\shortauthors{Torres}
\shorttitle{\estar}

\begin{document} 
\submitted{Accepted for publication in Monthly Notices of the Royal Astronomical Society}

\title{$\eta$ Geminorum: an eclipsing semiregular variable star orbited by a companion
surrounded by an extended disc}

\author{
Guillermo Torres\altaffilmark{1}, and
Kristy Sakano\altaffilmark{1,2}
}

\altaffiltext{1}{Center for Astrophysics $\vert$ Harvard \&
  Smithsonian, 60 Garden St., Cambridge, MA 02138, USA;
  gtorres@cfa.harvard.edu}
  
\altaffiltext{2}{Naval Air Warfare Center Aircraft Division,
22347 Cedar Point Road, Patuxent River, MD 20670, USA}

\begin{abstract}
We report 11~yr of spectroscopic monitoring of the M-type asymptotic giant branch star
\estar, a semiregular variable and a known spectroscopic
binary with a period of 8.2~yr. We combine
our radial velocities with others from the literature to provide an
improved spectroscopic orbital solution giving a period of 2979~days,
which we then use to predict past times of eclipse. We examine
archival photometry from amateur variable star observers, and other
sources, and find many instances of dimmings that occurred at the
right time. This confirms previous indications that the system is
eclipsing, and it now ranks among those
with the longest known periods. No secondary eclipses are
seen. The $\sim$0.4~mag eclipses lasting about 5 months are much too
deep to
be produced by a stellar companion. We propose instead that the
companion is surrounded by a large disk that is at least 1.5~au in
diameter, but is likely larger. We predict the center of the next eclipse will
occur on New Year's day, 2029.
\end{abstract}


\section{Introduction}
\label{sec:introduction}

$\eta$ Geminorum (HD~42995, HR~2216) is a late-type giant
\citep[M3.5\,Ib-II;][]{Abt:1985} and a well known semiregular variable
of type SRa, with a fairly consistent periodicity but with modest
changes in the amplitude and shape of the light curve. The photometric
period is about 230 days
\citep[see, e.g.,][]{Hoffmeister:1914, Vogelenzang:1928, vanSchewick:1950,
  Percy:1996, Percy:2001}. \estar\ has a visual companion approximately
  3~mag fainter discovered in 1881 by \cite{Burnham:1887}, which is
currently at a separation of 1.8 arcsec. The binary star designation
is WDS~J06149+2230. The orbital period has been estimated to be about
470~yr \citep{Baize:1980}. It is also a spectroscopic binary, which
makes the system a hierarchical triple.  The velocity changes were
first noticed by astronomers at the Lick Observatory in 1901 after the
third observation \citep{Campbell:1902}, and a provisional orbit was
published by \cite{Christie:1936}, although the reported period of 1875 days
was erroneous.  The correct period was established a few years later
in a study by \cite{McLaughlin:1944}, who derived an eccentric orbit
($e = 0.53$) with a period $P = 2983$ days, or 8.2 years.

In the same study those authors reported contemporaneous photometric
observations over about 13 years, and pointed out that the two deepest
minima, in 1931 and 1939, happened to be close to the times at which
primary eclipses would be expected from their spectroscopic orbit,
should the system be eclipsing.  However, given the very wide
separation of the components, they did not consider the coincidence to
be conclusive. If confirmed, \estar\ would rank among the longest
period eclipsing binaries known, even today.  \cite{vanSchewick:1950}
identified two other relatively deep minima in earlier brightness
measurements that seemed to fit the pattern suggested by
\cite{McLaughlin:1944}, and proceeded to infer a photometric period of
2984 days, essentially identical to the one from the spectroscopic
orbit. Neither period was given with an associated uncertainty.

After the 1950s, it seems that most studies dealing with other aspects
of \estar\ have generally {\it assumed\/} that the system is eclipsing,
although others \citep[e.g.,][]{Woolf:1973, Hassforther:2007} remained skeptical. Amateur variable
star observers have been more confident, organizing several
observing campaigns over the years hoping to record dips in brightness at the
expected times, with partial success.\footnote{See, e.g.,
  \cite{Bohme:1980, Bohme:1989}, and the following notes from the
  British Astronomical Association, Variable Star Section:
  \url{http://www.britastro.org/vss/The2004EclipseofEtaGeminorum.pdf},
  and \url{http://www.britastro.org/vss/EtaGem\_VSOTY2012.pdf}.}

To our knowledge there has been no definitive study of the persistence
of eclipses, although by now there are decades more of observations by
both amateur and professional variable star observers that are
potentially useful to that end.  This is therefore one of the main
goals of this paper. Additionally, there has been no modern redetermination
of the spectroscopic orbit of \estar\ after the one published nearly
80 years ago by \cite{McLaughlin:1944}. That
analysis relied on radial velocity measurements of considerably poorer
precision than today's instruments can deliver. Consequently, as a
second goal of this paper we monitored the object spectroscopically
over a period of 11.5 years (1.4 cycles of the spectroscopic orbit), aiming to confirm
and improve upon the previous orbit. Finally, it is of interest to
discuss in some detail the nature of the companion, in the context of
other eclipsing binaries of very long period such as the well known
systems $\epsilon$~Aur ($P = 27.1$~yr) and EE~Cep ($P = 5.6$~yr), or
the more recently discovered TYC-2505-672-1 system \citep[$P =
  69.1$~yr;][]{Rodriguez:2016, Lipunov:2016}. In all three of them,
and in a few others, the eclipsing object is surrounded by a large
opaque disc.  We will show that the evidence in \estar\ points to a
similar conclusion.

Our spectroscopic observations are presented in
Section~\ref{sec:spectroscopy}. In Section~\ref{sec:orbit} we combine our
radial velocities with other measurements from the literature to
derive an improved spectroscopic orbit for \estar. The extensive
photometric observations available for the object are discussed in
Section~\ref{sec:eclipses}, from which we compile the most complete
list of dimming events occurring near the times of primary eclipse, as
predicted from our updated spectroscopic orbit. These instances leave
no doubt about the eclipsing nature of \estar, and allow a substantial
improvement in the ephemeris of the system. The nature of the
secondary is discussed in Section~\ref{sec:discussion}. Final remarks
are given in Section~\ref{sec:conclusions}.

\section{Spectroscopic Observations}
\label{sec:spectroscopy}

\estar\ was placed on the observing program at the Center for
Astrophysics (CfA) as part of an effort to monitor about a dozen
semiregular variables. Observations began in September of 1993, and
continued until April of 2005 with two nearly identical instruments on
the 1.5m Wyeth reflector at the (now closed) Oak Ridge Observatory
(Massachusetts, USA) and the 1.5m Tillinghast reflector at the Fred
L.\ Whipple Observatory (Arizona, USA). These echelle instruments
\citep[Digital Speedometers;][]{Latham:1992} used intensified
photon-counting Reticon detectors that recorded a single order
45~\AA\ wide centered at a wavelength of 5187~\AA, featuring the
\ion{Mg}{1}\,b triplet. The resolving power was $R \approx 35,\!000$. We
gathered 137 spectra with the Wyeth reflector, and 6 with the
Tillinghast reflector. The signal-to-noise ratios (SNRs) range from 10
to 80 per resolution element of 8.5~\kms.  Wavelength solutions relied
on exposures of a thorium-argon lamp bracketing each science
exposure. The zero point of the velocity system was monitored with
observations of the sky early in the evening and in the morning, and
small run-to-run corrections based on them were applied as described
by \cite{Latham:1992}, and were usually smaller than 2~\kms. 

Our typical procedures for deriving radial velocities from similar
material involve cross-correlation against a synthetic template, using
the {\sc XCSAO} task running under {\sc IRAF}.\footnote{{\sc IRAF} is distributed
  by the National Optical Astronomy Observatories, which is operated
  by the Association of Universities for Research in Astronomy, Inc.,
  under contract with the National Science Foundation.} The late
spectral type of \estar\ is such that the spectrum is nearly saturated with
molecular features, which theoretical models still struggle to reproduce.
Although we had initially adopted a template from \cite{Husser:2013}
based on PHOENIX models, we found better results using instead a
high-SNR spectrum of the same star as the template.  The scatter from
our orbital solution described later was smaller, and the correlation
coefficients higher. To transfer these relative velocities to an
absolute frame of reference, we added an offset derived by
cross-correlating the observed template against the best-matching
synthetic PHOENIX spectrum with a temperature of $T_{\rm eff} =
3700$~K, $\log g = 0.0$, solar metallicity, and no rotational
broadening. While the transfer to the absolute frame may not be
perfect because the lines in the synthetic template do not exactly match those
in \estar, we are only concerned here with changes in the radial
velocities, so a small systematic offset is of no consequence for this
work. The velocities derived in this manner are presented in
Table~\ref{tab:rvs}, along with their formal uncertainties as returned
by {\sc XCSAO}.

\setlength{\tabcolsep}{6pt}  
\begin{deluxetable}{lccc}[!t]
\tablewidth{0pc}
\tablecaption{CfA Radial Velocities for \estar \label{tab:rvs}}
\tablehead{
\colhead{HJD} &
\colhead{$RV$} &
\colhead{$\sigma_{\rm RV}$} &
\colhead{Orbital Phase}
\\
\colhead{(2,400,000+)} &
\colhead{(\kms)} &
\colhead{(\kms)} &
\colhead{}
}
\startdata
  49260.9110  &  21.72  &  0.21  &  0.602 \\
  49284.8984  &  22.55  &  0.32  &  0.610 \\
  49289.8934  &  22.17  &  0.21  &  0.612 \\
  49295.9109  &  21.81  &  0.21  &  0.614 \\
  49313.8304  &  22.29  &  0.21  &  0.620 
\enddata
\tablecomments{Orbital phases were computed from the ephemeris given in
  Section~\ref{sec:orbit}. (The full version of this table
  is provided as supplementary material.)}
\end{deluxetable}
\setlength{\tabcolsep}{6pt}  

\section{Spectroscopic Orbit}
\label{sec:orbit}

Our main reason for attempting to improve upon the spectroscopic
orbital solution of \cite{McLaughlin:1944} is to use predictions for
the times of eclipse to identify dips in brightness in the photometric
observations going back a century or more, which might reasonably be
interpreted as eclipse events.  An initial orbital solution using our
velocities gave elements quite similar to those of
\cite{McLaughlin:1944}, with an uncertainty in the period of 9 days
and an uncertainty in the time of periastron passage of 23 days.
Projected backward to the earliest brightness measurements from the
1870s, the eclipse predictions are only good to about 140 days, a
consequence of the fact that our observations cover just slightly over
1.4 binary cycles. This can be improved significantly by considering
other sources of radial velocities together with ours.

Four small sets of historical measurements for \estar\ are available
between the years of 1900 and 1922: 11 velocities from the Lick
Observatory \citep{Campbell:1928}, 3 from the Cape Observatory
\citep{Lunt:1919} of which the first is clearly erroneous and was
rejected, 4 from the Mount Wilson Observatory \citep{Abt:1970}, and 2
from the Dominion Astrophysical Observatory \citep{Harper:1934}. To the
extent possible we transformed them all to the system of the Lick
measurements by applying small offsets published by \cite{Moore:1932}, and we
assigned them all uncertainties of 1~\kms. These four groups of
observations, and one additional measurement described below, were
considered as a single data set.  We also used the 142 velocities of
\cite{McLaughlin:1944} made between 1930 and 1941, with larger adopted
uncertainties of 2~\kms. For the next 50 years \estar\ appears to have
been almost completely neglected by spectroscopists: we have located
only a single radial velocity measurement reported by
\cite{Beavers:1986}, made at the Fick Observatory in 1979, which the
authors describe as being on the same system as the Lick
velocities. We assigned it an uncertainty of 1~\kms, and added it to
the first data set for simplicity. A more recent and extensive set of
129 high-quality measurements was reported by \cite{Eaton:2020}
covering the years 2003--2008. This partly overlaps with our own observations, and
extends the total spectroscopic coverage by three years.
Uncertainties for these measurements were taken to be 0.2~\kms.

A joint orbital fit using all radial velocities was carried out with a
Markov chain Monte Carlo (MCMC) procedure, using the {\sc
  emcee}\footnote{\url{https://emcee.readthedocs.io/en/stable/index.html}}
package of \cite{Foreman-Mackey:2013}. We used uniform priors over
suitable ranges for all jump variables. In addition to the standard
orbital elements $P$, $T_{\rm peri}$, $K_1$, and $\gamma$, we elected
to use eccentricity parameters $\sqrt{e}\cos\omega_1$ and
$\sqrt{e}\cos\omega_1$ instead of $e$ and $\omega_1$, where $\omega_1$
is the argument of periastron for the primary. Because there are four
separate data sets that may be offset from one another, we solved for
a different center-of-mass velocity for each one: $\gamma_{\rm old}$,
$\gamma_{\rm McL}$, $\gamma_{\rm CfA}$, and $\gamma_{\rm Eat}$.
Additionally, we solved for four radial velocity ``jitter'' terms
$\sigma_{\rm j,old}$, $\sigma_{\rm j,McL}$, $\sigma_{\rm j,CfA}$, and
$\sigma_{\rm j,Eat}$, to be added quadratically to the internal errors
in order to account not only for the possibility that errors may be
underestimated, but also for the extra scatter that may be caused by
the $\sim$230-day pulsations in \estar.

\begin{figure*}
\epsscale{1.18}
\plotone{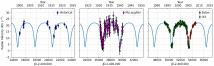}

\figcaption{Radial-velocity measurements for \estar\ from various
  sources, as labelled. The isolated 1979 measurement of \cite{Beavers:1986}
  that we have included as part of the ``historical'' data set is
  not shown, for clarity. The solid curve is our best-fitted model
  described in Section~\ref{sec:orbit} that incorporates the eclipse
  timings. \label{fig:rvs}}
\end{figure*}

The results are given in the second column of Table~\ref{tab:results} (Solution~1),
together with derived properties listed at the bottom. The benefit of
adding the other data sets to ours is seen in the significant reduction
in the uncertainties for the period (a factor of 9) and time of
periastron passage (a factor of 4). Predictions for the times of
eclipse are about a factor of 7 better than before. All radial
velocity observations are shown with our best-fitted model in
Figure~\ref{fig:rvs}.

\setlength{\tabcolsep}{1pt}
\begin{deluxetable*}{lcc}
\tablewidth{0pc}
\tablecaption{Orbital Solutions for \estar \label{tab:results}}
\tablehead{
\colhead{Parameter} &
\colhead{Solution 1} &
\colhead{Solution 2}
}
\startdata
 $P$ (day)                    & $2978.4 \pm 1.2$\phn\phn\phn  & $2978.90 \pm 0.49$\phm{222}           \\ 
 $T_{\rm peri}$ (HJD$-$2,400,000) & $44488.4 \pm 5.5$\phm{2222}  & $44489.1 \pm 4.5$\phm{2222}  \\ 
 $\sqrt{e}\cos\omega_1$       & $-0.7351 \pm 0.0044$\phs      & $-0.7382 \pm 0.0030$\phs         \\
 $\sqrt{e}\sin\omega_1$       & $+0.076 \pm 0.011$\phs        & $+0.075 \pm 0.011$\phs           \\
 $K_1$ (\kms)                 & $9.417 \pm 0.086$             & $9.438 \pm 0.084$            \\
 $\gamma_{\rm CfA}$ (\kms)    & $+17.597 \pm 0.062$\phn\phs   & $+17.601 \pm 0.063$\phn\phs  \\ 
 $\gamma_{\rm old}$ (\kms)    & $+18.96 \pm 0.30$\phn\phs     & $+18.93 \pm 0.28$\phn\phs    \\ 
 $\gamma_{\rm McL}$ (\kms)    & $+17.43 \pm 0.25$\phn\phs     & $+17.42 \pm 0.25$\phn\phs    \\ 
 $\gamma_{\rm Eat}$ (\kms)    & $+17.84 \pm 0.11$\phn\phs     & $+17.791 \pm 0.097$\phn\phs  \\ 
 $\sigma_{\rm j,CfA}$ (\kms)  & $0.683 \pm 0.052$             & $0.679 \pm 0.051$            \\
 $\sigma_{\rm j,old}$ (\kms)  & $0.74\pm 0.39$                & $0.71 \pm 0.38$              \\
 $\sigma_{\rm j,McL}$ (\kms)  & $2.05 \pm 0.25$               & $2.05 \pm 0.25$              \\
 $\sigma_{\rm j,Eat}$ (\kms)  & $0.924 \pm 0.065$             & $0.929 \pm 0.066$            \\
  $f_{\rm tim}$                & \nodata                       & $1.24 \pm 0.26$             \\ [1ex]
\hline \\ [-1.5ex]
\multicolumn{3}{c}{Derived quantities} \\ [0.5ex]
\hline \\ [-1.5ex]
 $e$                          & $0.5463 \pm 0.0061$           & $0.5507 \pm 0.0041$          \\
 $\omega_1$ (deg)             & $174.13 \pm 0.85$\phn\phn     & $174.20 \pm 0.82$\phn\phn    \\
 $f(M)$ ($M_{\sun}$)          & $0.1515 \pm 0.0044$           & $0.1509 \pm 0.0042$          \\
 $M_2 \sin i / (M_1 + M_2)^{2/3}$ ($M_{\sun}$)  & $0.5330 \pm 0.0052$ & $0.5324 \pm 0.0049$  \\
 $a_1 \sin i$ ($10^6$ km)     & $323.0 \pm 3.2$\phn\phn       & $322.7 \pm 3.0$\phn\phn      
\enddata
\tablecomments{Solution~1 uses only radial velocities. Solution~2 adds
  the eclipse timings from Table~\ref{tab:timings}. The $\sigma_{\rm j}$
  symbols are jitter terms added quadratically to the internal velocity
  errors, and $f_{\rm tim}$
  represents a scale factor for the eclipse timing measurements. Posterior
  distributions for the derived quantities are constructed from those
  of the quantities involved.}
\end{deluxetable*}
\setlength{\tabcolsep}{6pt}

\section{Eclipses}
\label{sec:eclipses}

\estar\ has been a target of variable star observers for a
century and a half. Several organizations over the world have systematically
collected those observations, made mostly by amateur astronomers, and
in some cases they have made efforts to homogenize them. Here we consider
four large sets of brightness measurements from the American
Association of Variable Star Observers (AAVSO, $\sim$28,200
measurements),\footnote{\url{https://www.aavso.org/}} the Association
Fran\c{c}aise des Observateurs d'\'Etoiles Variables (AFOEV, 
$\sim$10,800 measurements) [French Association of Variable-Star
  Observers],\footnote{\url{https://cdsarc.u-strasbg.fr/ftp/pub/afoev/}}
the British Astronomical Association--Variable Star Section (BAA-VSS,
$\sim$6,400 measurements),\footnote{\url{https://britastro.org/photdb/data.php}}
and the Variable Star Observers League in Japan (VSOLJ, 
$\sim$6,100 measurements),\footnote{\url{http://vsolj.cetus-net.org/database.html}} all
with data as of June of 2022. We have used only observations
in or near the visual band, which are the most common. In addition we
have also examined shorter sets of data by \cite{McLaughlin:1944},
\cite{Percy:2001}, and from the Czech Astronomical Society--Variable
Star and Exoplanet Section
(CAS-VSES).\footnote{\url{http://var2.astro.cz/EN/}} When possible,
for the largest data sets we have separated the measurements provided
with one decimal place (mostly made visually) from the more precise
ones given to two or more decimal places. This is to guard against the
possibility of slight differences in the magnitude scales. The format
provided for the BAA-VSS observations did not allow this distinction.
In most of these data sets, but especially among the more precise
data, brightness fluctuations due to the $\sim$230-day semiregular
pulsations are obvious and have total amplitudes of up to 0.5~mag in
the visual \citep[see, e.g.,][]{Cristian:1995, Percy:1996, Percy:2001,
  Hassforther:2007}.

\begin{figure*}
\epsscale{1.18}
\plotone{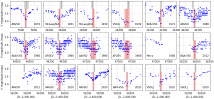}

\figcaption{Primary eclipses of \estar\ recorded in archival
  photometric measurements since 1874, as labelled. The gray shaded
  areas in the top third of each panel mark the predicted eclipse
  times from our Solution~1 in Table~\ref{tab:results}, with their
  uncertainty. The red shaded areas in the bottom two thirds
  represent the times of minimum estimated by eye, with their
  uncertainties. These are the timing measurements incorporated into
  Solution~2 (along with two others; see the text).  Two of the panels
  with no gridding and no lower shaded area show dimmings that happened
  approximately at the expected times, but were considered to be too
  poorly defined to measure and use.\label{fig:eclipses}}
\end{figure*}

Armed with the improved spectroscopic orbital elements from the
previous section, we examined these sources of photometric
observations and identified many instances where light minima were
close to the predicted times of eclipse (see
Figure~\ref{fig:eclipses}).  This leaves no doubt as to the eclipsing
nature of \estar. In all cases the events are primary eclipses; no
compelling case of a secondary eclipse was found (secondary minima are expected about
504 days later than primary eclipse, though with somewhat larger
uncertainties).  We note that
because of the intrinsic variability of the star, the primary minima vary in
depth depending on whether or not the drop in brightness from the
eclipse is superimposed with a change due to the semiregular
variations. The eclipse centers may also be affected for the same
reason. The gray shaded areas in the top third of each panel in
Figure~\ref{fig:eclipses} mark the predicted times of eclipse and
their formal uncertainties.

Eclipse timings provide a much stronger constraint on the ephemeris of
a binary than radial velocities.\footnote{We note in passing that the AAVSO
  observations contain a valuable subset of measurements from the
  1870s originally made by Edward Schoenfeld (1828--1891), and
  published by \cite{Valentiner:1900}. These are the oldest publicly
  available brightness measurements for \estar. They were later
  reprocessed by \cite{Burnashev:2009} using modern magnitudes for the
  comparison stars, and as a result the precision is now considerably
  improved. These observations happen to be the first to have recorded
  an eclipse of \estar, in January of 1874, which therefore carries
  the highest weight for improving the ephemeris. This important event
  seems to have been overlooked by previous investigators, even though
  the observations are contained in the public AAVSO archive.} To take
advantage of this information, we measured the center of each event in
Figure~\ref{fig:eclipses} by eye, and assigned realistic
uncertainties.  These determinations are shown with red shaded areas
in the lower two thirds of each panel, and are listed in
Table~\ref{tab:timings}. Two additional measurements were made
graphically from series of observations in 1980 and 1988 shown in
figures by \cite{Bohme:1980, Bohme:1989}, and were added to the
table. We have not included four additional timings reported
  by \cite{vanSchewick:1950} because they are given without
  uncertainties, and we do not have access to the original
  observations. The dates of those events are JD $2,\!411,\!452$,
  $2,\!423,\!387$, $2,\!426,\!360$, and $2,\!426,\!375$.

\setlength{\tabcolsep}{4pt}
\begin{deluxetable*}{rccccl}
\tablewidth{0pc}
\tablecaption{Primary Eclipse Timing Measurements for \estar \label{tab:timings}}
\tablehead{
\colhead{Cycle} &
\colhead{Year} &
\colhead{Time of Eclipse} &
\colhead{$\sigma$} &
\colhead{$O-C$} &
\colhead{Source}
\\
\colhead{} &
\colhead{} &
\colhead{(JD$-2,\!400,\!000$)} &
\colhead{(day)} &
\colhead{(day)} &
\colhead{}
}
\startdata
 $-$13 &  1874   &\phn5537  &   8 &  \phn$-$2\phs  &  AAVSO (two decimals)   \\
  $-$6 &  1931   &   26379  &   8 &  $-$13\phs     &  \cite{McLaughlin:1944} \\
  $-$5 &  1939   &   29376  &  10 &  \phn+5\phs    &  \cite{McLaughlin:1944} \\
  $-$5 &  1939   &   29360  &  20 &  $-$11\phs     &  VSOJL (one decimal)    \\
  $-$1 &  1971   &   41281  &   7 &  \phn$-$5\phs  &  BAA-VSS                \\
     0 &  1980   &   44274  &   7 &  \phn+9\phs    &  AAVSO (one decimal)    \\
     0 &  1980   &   44271  &  10 &  \phn+6\phs    &  BAA-VSS                \\
     0 &  1980   &   44272  &   6 &  \phn+7\phs    &  AFOEV (one decimal)    \\
     0 &  1980   &   44280  &  14 &  +15\phs       &  \cite{Bohme:1980}      \\
     1 &  1988   &   47253  &  20 &  \phn+9\phs    &  AAVSO (one decimal)    \\
     1 &  1988   &   47264  &  20 &  +20\phs       &  AFOEV (one decimal)    \\
     1 &  1988   &   47253  &   7 &  \phn+9\phs    &  \cite{Bohme:1989}      \\
     4 &  2012   &   56185  &  10 &  \phn+4\phs    &  BAA-VSS                \\
     5 &  2020   &   59160  &  10 &  \phn0         &  AAVSO (one decimal)    \\
     5 &  2020   &   59146  &   7 &  $-$14\phs     &  AAVSO (two decimals)   \\
     5 &  2020   &   59174  &   6 &  +14\phs       &  BAA-VSS                \\
     5 &  2020   &   59156  &   8 &  \phn$-$4\phs  &  VSOJL (one decimal)    \\
     5 &  2020   &   59147  &   7 &  $-$13\phs     &  VSOJL (two decimals)
\enddata

\tablecomments{Cycle numbers in the first column are counted from the
  primary eclipse nearest to the reference time of periastron passage
  in Table~\ref{tab:results}. That eclipse is at JD $2,\!444,\!265.2$. ``One
  decimal'' or ``two decimals'' in the last column refer to the formal precision with
  which the magnitudes are given in each data set (see the text).
  Timings from \cite{Bohme:1980, Bohme:1989} were determined by eye
  from figures presented in those reports.}
\end{deluxetable*}

We then performed a final orbital solution in which we combined the
radial velocity measurements with the eclipse timings in
Table~\ref{tab:timings}. A scale factor $f_{\rm tim}$ for the formal timing
errors was included as an additional adjustable parameter, with a
log-uniform prior. The results are shown in the last column of
Table~\ref{tab:results} (Solution~2). The precision of the period is improved by
more than a factor of 2 compared to the previous fit. Residuals for the timings are listed in
Table~\ref{tab:timings}.

\section{Discussion}
\label{sec:discussion}

Knowledge of the spectroscopic orbit of \estar\ provides information
about the mass of the companion. If we assume an edge-on orientation,
which must be very close to the truth given the wide separation, then
for a given adopted primary mass the companion mass is $M_2 = 0.5324
(M_1 + M_2)^{2/3}$, where the coefficient is that listed in
Table~\ref{tab:results}. To estimate $M_1$ we relied on other
properties measured for the star, as follows.

\cite{Baines:2021} used long-baseline interferometry to measure the
angular diameter of \estar, and reported a value of $\theta_{\rm LD} =
12.112 \pm 0.024$~mas corrected for limb darkening. This is consistent
with several other determinations \citep[see,
  e.g.,][]{Mozurkewich:2003, Richichi:2003, Mondal:2005}.
\cite{Baines:2021} converted this to an absolute radius using a
distance inferred from the parallax of the star ($\pi_{\rm DR2} = 4.73
\pm 1.02$~mas), which they adopted from the second data release (DR2)
of \gaia\ \citep[][source identifier
  3377072212925223424]{GaiaDR2:2018}.\footnote{Parallax determinations
  for \estar\ have been problematic: the original \hip\ catalog
  \citep[][source identifier HIP~29655]{ESA:1997} reported a much different value of
  $\pi = 9.34 \pm 1.99$~mas. The revised \hip\ catalog
  \citep{vanLeeuwen:2007} gives $\pi = 8.48 \pm 1.23$~mas, and the
  most recent release of \gaia\ \citep[DR3;][source ID
    3377072212924335488]{GaiaDR3:2022} gives no value at all. 
    Because of the large angular size of \estar\ and its pulsating
    nature, changes in brightness across its surface have the potential
    to alter the center of light, affecting the parallax determinations
    \citep[e.g.,][]{McDonald:2012}.
    We accept the parallax choice of \cite{Baines:2021}.} They obtained $R_1 = 275
\pm 76~R_{\sun}$, where we have adopted the larger of their asymmetric
errors. They also performed a fit to the spectral energy distribution
of the star to estimate the bolometric flux, $F_{\rm bol}$. From the
angular diameter and flux they derived an effective temperature of
$T_{\rm eff} = 3502 \pm 30$~K. Their inferred luminosity for the star is
then $L_1 = 10,\!276 \pm 4,\!445~L_{\sun}$.

\begin{figure}
\epsscale{1.18}
\plotone{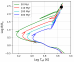}

\figcaption{Measured radius and temperature of \estar\ according to
\cite{Baines:2021}, shown against solar-metallicity model isochrones
from \cite{Choi:2016} for four different ages, as labelled. The 110~Myr
model presents the best match to the observations. The cross
marks the expected location of the spectroscopic secondary on that
isochrone, based on its calculated mass.\label{fig:mist}}
\end{figure}

In Figure~\ref{fig:mist} we show the above radius and temperature of
the star against a representative set of model isochrones from the
MIST series of \cite{Choi:2016}. The location of \estar\ is near the
tip of the asymptotic giant branch. Based on these models we infer a
primary mass of $M_1 \approx 5.1~M_{\sun}$ at an age of about 110~Myr.
This then leads to a secondary mass of $M_2 \approx 2.0~M_{\sun}$
corresponding to an early A-type star ($T_{\rm eff} \approx 9000$~K),
provided it is a single star. The luminosity would be
$L_2 \approx 16~L_{\sun}$, nominally some 600 times less luminous than the primary.
Such a star would have a radius of $R_2 \approx 1.65~R_{\sun}$ according
to the models, which is far too small to produce an eclipse of the depth
shown in Figure~\ref{fig:eclipses}. The inescapable
conclusion is that the secondary must be surrounded by a much larger
and at least partially opaque structure, which we speculate may be a
disc.

\estar\ then appears to be another example of a class of wide binaries
in which the primary is eclipsed by a disc around the companion. The
prototype of this small and diverse class of objects is the much
studied 27.1-yr binary $\epsilon$~Aur, an F-type supergiant.  The
large tilted disk enshrouding the B-type companion of $\epsilon$~Aur has been estimated
to have a radius of 4--6~au \citep[see, e.g.,][]{Hoard:2010,
  Mourard:2012}, and causes eclipses that last about 2 years. A disc
of similar size (roughly 2--7~au) has been proposed to be surrounding
the hot (stripped red giant) pre-He white dwarf companion of the
M-type giant primary in TYC-2505-672-1, currently the eclipsing binary
with the longest orbital period \citep[$P = 67.1$~yr;][]{Lipunov:2016,
  Rodriguez:2016}. In this case the eclipses last for an astonishing
$\sim$3.5 years.  The shorter-period eclipsing binary EE~Cep ($P =
5.6$~yr) has an invisible companion to the B5 giant primary embedded
in a disc with a roughly 2~au diameter \citep{Pienkowski:2020}. The
eclipses are highly variable in depth, and can last up to 60 days.

A disc such as the one we propose in \estar\ can potentially
contribute some amount of infrared excess to the system, which
could provide information on its nature and composition. The spectral energy
distribution of the star was examined by \cite{McDonald:2012, McDonald:2017} based on
brightness measurements between 0.42 and 25~$\mu$m, as part of
much larger samples of \hip\ and {\it Tycho-2\/} sources. They
reported finding no significant excess between 4.3 and 25~$\mu$m
within the photometric uncertainties. Studies of this kind are made
more challenging in \estar\ because of the intrinsic variability, and
to some extent also because of the presence of the visual companion
that may not always be resolved in the photometric measurements.

\begin{figure}
\epsscale{1.18}
\plotone{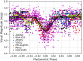}

\figcaption{Photometry for \estar\ near the primary eclipse,
  phase-folded with the ephemeris from Solution~2 in
  Table~\ref{tab:results}. Only data reported with two decimal places
  are displayed, where that distinction can be made, and we have
  attempted to align the measurements from the different sets by
  shifting them vertically to the same median value as the AAVSO set.
  The solid curve represents our best-fitted {\sc JKTEBOP} eclipsing
  binary model described in the text. Data outside of eclipse were
  downweighted to prevent the scatter from the semiregular variations
  from overwhelming the $\chi^2$ of the
  solution.\label{fig:phasefolded}}
\end{figure}

To gain some understanding of the dimensions of the disc in \estar, we
use the shape of the eclipse as determined from the observations.
Figure~\ref{fig:phasefolded} shows a phase-folded plot of the
photometric data near the primary eclipse, based on the best ephemeris
from Section~\ref{sec:orbit}.  The eclipse is roughly 0.4~mag deep,
although there is significant variation in the depth of individual
events caused by the pulsations, as seen earlier in
Figure~\ref{fig:eclipses}. This depth implies a flux decrement of
about 30\%. To first order this would be the ratio of the surface area
of the occulting object relative to that of the primary. The total
duration of the eclipse is about 5 months, although the precise
contact points are difficult to establish given the scatter in the
photometry.  A binary model fitted to these data is indicated with the
solid curve in Figure~\ref{fig:phasefolded}. The fit was performed
using the {\sc
  JKTEBOP}\footnote{\url{https://www.astro.keele.ac.uk/jkt/codes/jktebop.html}}
code of \cite{Southworth:2004}, with linear limb-darkening and gravity
darkening coefficients appropriate for the primary star, and assuming
that the secondary contributes no significant light given the lack of
secondary eclipses. The ephemeris and eccentricity parameters were
held fixed at the values in Table~\ref{tab:results}, and the mass
ratio was set from the mass estimates given earlier, although the
results are insensitive to this quantity.

By design {\sc JKTEBOP} assumes the companion is a star, which will
have an essentially circular profile, whereas a disc might be expected to
be a flattened structure. Nevertheless, we may view the results
from our fit as a limiting case of a companion that is a completely opaque and circular
disc seen face-on, regardless of whether or not this is a realistic
orientation. Light curve fits such as this have no absolute scale, and
only yield relative quantities. The relative radius of the secondary returned
by the model is $R_2/a \approx 0.10$, where $a$ is the semimajor axis
of the binary. Experiments show that this value appears quite robust against changes in limb
darkening or gravity darkening.  With the adopted masses and the
orbital period, we derive a semimajor axis of $a \approx
1670~R_{\sun}$, which leads to a diameter for the face-on disc of
$\sim$330~$R_{\sun}$, or about 1.5~au.  This is strictly a lower limit
to its size, as any other orientation would lead to a smaller
projected area, and the requirement to still block 30\% of the surface
of the primary means the disc diameter would have to be larger.
Furthermore, the disc is not expected to be completely opaque, so
meeting the same requirement with any radially-decreasing density
distribution would imply an even larger size. We do not attempt
here to perform a more sophisticated modelling of the disc properties or a full exploration
of possible geometries (orientations, aspect ratio, etc.), as
this likely requires better constraints on the eclipse shape
than the current observations can provide.

The formal impact parameter from our light curve fit is $\sim$0.60
(inclination angle $\sim$81\arcdeg), although we do not attach a high
significance to this result given the tentative nature of the
solution. On the other hand, the more firmly determined relative
radius of the primary star is $R_1/a \approx 0.17$. Interestingly,
converting back to an absolute scale results in $R_1 \approx
280~R_{\sun}$, which happens to be very close to the primary radius
derived from its measured angular diameter and the parallax
\citep[$R_1 = 275~R_{\sun}$;][]{Baines:2021}. 

Finally, the linear semimajor axis of the spectroscopic orbit
corresponds to approximately 7.8~au, which translates to an angular
semimajor axis of about 37~mas at the adopted distance. Projected
angular separations can reach 57~mas (and will do so in mid 2025),
which should be easily accessible by various astrometric techniques.
Lunar occultation observations of \estar\ that have been used in the past to
measure its angular diameter \citep{Richichi:2003, Mondal:2005} have
not revealed any sign of the spectroscopic companion.  The 110~Myr
MIST isochrone employed earlier predicts a magnitude contrast of 5~mag in
$V$ (assuming the secondary is a single star), but much less favorable
values in the near infrared ($\Delta H \approx 9.3$~mag, $\Delta K
\approx 9.9$~mag). This assumes no obscuration of the secondary by its
surrounding disc.

\section{Conclusions}
\label{sec:conclusions}

We have used our new radial velocity measurements combined with
others from the literature to update the 2979-day (8.2 yr)
spectroscopic orbit of the M3.5\,Ib-II semiregular variable \estar.
Inspection of nearly 150 years of archival photometry collected by
several associations of variable star observers around the world, as well as by
professional astronomers, have allowed us to identify a number of
instances in which clear dimmings occur very near the times of primary
eclipse predicted from our spectroscopic orbit.  Eighteen independent
detections from different data sets for 8 separate eclipses have been
compiled, and remove any doubt that primary eclipses recur faithfully
every 8.2 yr. No secondary eclipses are seen.

Combining the measured times of eclipse with the radial velocities
yields a much improved ephemeris, with which we predict the center of
the next eclipse will occur on January 1st 2029 (JD $2,\!462,\!138.6 \pm
3.3$).

\estar\ now ranks among the longest-period eclipsing binaries that
have been confirmed to date. Only ten examples were previously known
with periods longer than 5~yr, not all of which have known orbits
(spectroscopic or astrometric). \estar\ is only the sixth case with an
orbit. A listing of all such long-period eclipsing systems of which we
are aware is presented in Table~\ref{tab:longperiod}, with an
indication of the variability type, spectral type, and which have reliable
orbits determined. References are provided for the orbital determinations.
Several other candidate long-period systems have been
proposed, but require confirmation; they are not included in the table.

\setlength{\tabcolsep}{6pt}
\begin{deluxetable*}{lcclll}
\tablewidth{0pc}
\tablecaption{Confirmed Eclipsing Binaries with Orbital Periods Longer than 5 yr\label{tab:longperiod}}
\tablehead{
\colhead{Name} &
\colhead{Period (yr)} &
\colhead{Orbit} &
\colhead{Variability Type} &
\colhead{Spectral Type} &
\colhead{Orbit reference}
}
\startdata
 TYC-2505-672-1   &  69.07   &   No   &  \nodata     &  M1III+sdB         & \nodata \\
 $\epsilon$ Aur   &  27.09   &  Yes   &  EA/GS       &  A9Ia              & \cite{Stefanik:2010} \\
 VV Cep           &  20.34   &  Yes   &  EA/GS+SRC   &  M2epIa-Iab+B8:eV  & \cite{Wright:1977} \\
 V381 Sco         &  17.92   &   No   &  EA/GS+SRC   &  A8II              & \nodata \\
 $\gamma$ Per     &  14.64   &  Yes   &  EA/GS       &  G9III+A2-III:     & \cite{Pourbaix:2000} \\
 V383 Sco         &  13.35   &   No   &  EA/GS+SRC   &  F0Iab:e           & \nodata \\
 31 Cyg           &  10.36   &  Yes   &  EA/GS/D     &  K3Ib+B2IV-V       & \cite{Videla:2022} \\
 AZ Cas           &   9.32   &  Yes   &  EA          &  K5Iab-Ib          & \cite{Ashbrook:1956} \\
 $\eta$~Gem       & 8.15     &  Yes   & EA/GS+SRA    &   M3.5Ib-II        & This paper \\
 KIC 5273762      &   7.33   &   No   &  EA/GS+SR    &  gM3.0             & \nodata \\
 EE Cep           &   5.61   &   No   &  EA          &  B5:nev            & \nodata
\enddata
\tablecomments{The variability type has been extracted from the General Catalog
of Variable Stars \citep[GCVS;][]{Samus:2017}, or assembled from other sources in the literature. The
GS notation refers to a system in which one or both components are giants or
supergiants. SRA and SRC are two subtypes of semiregular variables defined
in the GCVS. Spectral types are from SIMBAD or other sources in the literature.
The $\gamma$~Per binary has both a spectroscopic and an
  astrometric orbit. For KIC~5273762 we use this shorter designation
  in the table; the designation in the discovery paper is
  ASASSN-V~J192543.72+402619.0 \citep{Jayasinghe:2018}.}
\end{deluxetable*}
\setlength{\tabcolsep}{6pt}

The primary eclipses of \estar\ are about 0.4~mag in depth, on
average, and last for approximately 5 months.  Given the large size of
the primary star \citep[275~$R_{\sun}$;][]{Baines:2021}, the eclipses
are far too deep to be produced by the spectroscopic companion. We
propose they are instead caused by an extended disc surrounding the
secondary, similar to the discs found in several other long-period
eclipsing binaries with evolved primary components, such as
$\epsilon$~Aur. The properties of the disc around \estar\ are not well
established at present, although simple arguments suggest it is at
least 1.5~au in diameter, and very likely larger. The difficulties are
partly due to confusion stemming from the semiregular variations that
add scatter to the light curve, and that have amplitudes similar to
the depth of the eclipse.  Further improvements in our knowledge may
come from the next opportunity for study that will occur near the
beginning of 2029. That eclipse will be well placed for
observation, unlike others that have occurred with \estar\ too close to
the Sun.

\begin{acknowledgements}

The spectroscopic observations of \estar\ at the CfA were obtained
with the assistance of J.\ Caruso, D.\ W.\ Latham, R.\ P.\ Stefanik,
and J.\ Zajac. We thank R.\ J.\ Davis for maintaining the database of
echelle spectra.  We are also grateful to M.\ McEachern (Wolbach
Library) for her valuable help in locating and providing copies of
some of the historical papers for \estar.
Support is also acknowledged from the Smithsonian Minority Awards Program, run by the
Smithsonian Institution Office of Fellowships (Washington, DC).
The anonymous referee is thanked for helpful suggestions.
We acknowledge with thanks the variable star observations from the
AAVSO International Database contributed by observers worldwide and
used in this research.  The research has also made use of the AFOEV
database, operated at CDS, France, of the BAA Photometry Database
(UK), of the database of the Variable Star Observers League in Japan,
and of the database from the Variable Star and Exoplanet Section of
the Czech Astronomical Society.
The research has made use as well
of the SIMBAD and VizieR databases, operated at the CDS, Strasbourg,
France, and of NASA's Astrophysics Data System Abstract Service.

\end{acknowledgements}

\section{Data Availability}

The data underlying this article are available in the article and in
its online supplementary material.

\end{document}